\documentclass[12pt,a4paper]{article}
\usepackage[T2A]{fontenc}
\usepackage[utf8]{inputenc}
\usepackage[english]{babel}
\usepackage{amssymb}
\usepackage{amsmath}
\usepackage{graphicx}
\usepackage[small,sc,center]{titlesec}
\usepackage{indentfirst}
\usepackage{siunitx}
\usepackage[labelsep=period]{caption}
\usepackage{caption}

\begin{document}

\begin{center}
	\textbf{Theoretical mass estimates for the Mira--type variable R Hydrae}
	
	\vskip 3mm
	\textbf{Yu. A. Fadeyev\footnote{E--mail: fadeyev@inasan.ru}}
	
	\textit{Institute of Astronomy, Russian Academy of Sciences,
		Pyatnitskaya ul. 48, Moscow, 119017 Russia} \\
	
	Received April 4, 2023; revised April 27, 2023; accepted April 27, 2023
\end{center}

\textbf{Abstract} ---
Calculations of stellar evolution at initial abundances of helium $Y=0.28$ and heavier elements
$Z=0.014$ were done for stars with masses on the main sequence $1.7M_\odot\le M_\textrm{ZAMS}\le 5.2M_\odot$.
Evolutionary sequences corresponding to the AGB stage were used for modelling the pulsation period
decrease observed for almost two centuries in the Mira--type variable R~Hya.
Diminution of the period from $\Pi\approx$ 495 d in the second half of the eighteenth century to
$\Pi\approx 380$ d in the 1950s is due stellar radius decrease accompanying dissipation of the
radiation--diffusion wave generated by the helium flash.
For all the history of its observations R~Hya was the fundamental mode pulsator.
The best agreement with observations is obtained for eight evolutionary models with initial mass
$M_\textrm{ZAMS}=4.8M_\odot$ and the mass loss rate parameter of the Bl\"ocker formula $0.03\le\eta_\mathrm{B}\le 0.07$.
Theoretical mass estimates of R~Hya are in the range $4.44M_\odot\le M\le 4.63M_\odot$, whereas the
mean stellar radius ($421R_\odot\le \bar R \le 445R_\odot$) corresponding to the pulsation period
$\Pi\approx 380$ agrees well with measurements of the angular diameter by methods of the optical
interferometric imaging.

Keywords: \textit{stellar evolution; stellar pulsation; stars: variable and peculiar}

\section*{introduction}

The variable star R~Hya belongs to the long--period Mira--type pulsating variables (Samus' et al. 2017)
and its first observations were made by Jan Heweliusz in 1662 and Geminiano Montanari in 1670
(Hoffleit 1997; Zijlstra et al. 2002).
However systematic observations of R~Hya were conducted only since the second half of the XIXth century.
A particular attention to this star was paid after detection of a rapid decrease in the period
of its light variations
(Schmidt 1865; Chandler 1882; Gould 1882; Cannon and Pickering 1909; Ludendorff 1916; Nielsen 1926;
M\"uller 1929).
The most comprehensive review of R~Hya period change is given by Zijlstra et al. (2002).
According to these data the onset of R~Hya period decrease corresponds to the second half
of the XVIIIth century when the star oscillated with period $\Pi\approx 495$ d.
Since the end of the XVIIIth century the period decreased nearly linearly with the rate
$\dot\Pi\approx -0.58$ d/yr and the period decline ceased in $\approx 1950$ when the star oscillated
with period $\Pi\approx 380$ d.

Detection of absorption lines of the unstable radioactive element technetium with a half--life
$\tau\lesssim 2\times10^5$~yr in the spectrum of R~Hya
(Orlov and Shavrina 1984; Little et al. 1987; Lebzelter and Hron 2003)
gives a good indication that this late--type pulsating variable is the asymptotic giant branch (AGB)
star undergoing the thermal flash of the helium burning shell accompanied by the change of the surface
chemical composition due to the third dredge--up.
At the same time one should note that R~Hya is in the early TP--AGB stage since this star is
classified as the oxygen--rich Mira--type variable (Merrill 1946, 1957; Maehara 1971).
The higher abundance of oxygen in comparison with carbon is also demonstrated by observations of maser
emission of molecules OH (Lewis et al. 1995), $\textrm{H}_2\textrm{O}$ (Takaba et al. 2001) and
SiO (Humphreys et al. 1997).

Rough estimates based on evolutionary computations (Wood and Zarro 1981) show that the pulsation period
change observed in R~Hya is due to decrease of the stellar radius and luminosity after the maximum
of the thermal flash in the helium burning shell.
Unfortunately, the mass of the Mira--type variable R~Hya remains still uncertain because a thorough
analysis based on results of nonlinear stellar oscillations has not been done yet.

At present R~Hya is the only Mira--type variable with the known duration of the period decrease
($170~\textrm{yr}\lesssim \Delta t\lesssim 200~\textrm{yr}$)
as well as with the known values of the period at the onset ($\Pi_a^\star\approx 495$ d) and
at the end ($\Pi_b^\star\approx 380$ d) of this interval (Zijlstra, 2002).
In our previous paper (Fadeyev 2022) we have shown that analysis of the secular period change during
the thermal flash of the Mira--type variable T~UMi allowed us to obtain the reliable estimate of the stellar
mass using the consistent evolutionary and nonlinear stellar pulsation calculations.
Below we describe the results of similar calculations aimed at evaluating the mass $M$ and the radius $R$
of R~Hya.
The models of R~Hya computed in the present study are tested by comparison of theoretical estimates of the
mean stellar radius $\bar R$ with results of angular stellar diameter measurements obtained by methods of
the optical interferometry (Haniff et al. 1995; Ireland 2004; Woodruff et al. 2008).

\section*{evolutionary sequences of agb stars}

The results presented below are based on evolutionary computations of stars from the main sequence
up to the end of the AGB stage.
We considered the stars with initial masses $1.7M_\odot\le M_\textrm{ZAMS}\le 5.2M_\odot$ and the initial helium
abundance $Y=0.28$.
The initial metallicity (i.e. the abundance of elements heavier than helium) was assumed
to be the same as the solar metallicity $Z=0.014$ (Asplund 2009).

The evolutionary sequences were calculated with the program MESA version r15140 (Paxton et al. 2019).
Convective mixing was treated according to B\"ohm--Vitense (1958) with mixing length to pressure scale
height ratio $\alpha_\mathrm{MLT}=1.8$.
Additional mixing at boundaries of convection zone was taken into account by using the
prescription of Herwig (2000) with values of the overshooting parameter $f_\mathrm{ov}$ recommended
by Pignatari et al. (2016).
In particular, evolutionary stages before AGB were computed with $f_\mathrm{ov} = 0.014$ whereas during
the AGB phase the overshooting parameter at the inner boundary of the outer convection zone
was assumed to be $f_\mathrm{ov} = 0.126$.
Assumption on extended overshooting at the bottom of the outer convection zone provides a better agreement
with carbon and oxygen abundances observed in AGB stars (Herwig et al. 2003; Pignatari et al. 2016).
The rates of nuclear reactions and nucleosynthesis were calculated using the JINA Reaclib data base
(Cyburt et al. 2010).

Evolutionary phases prior to the AGB (i.e. when the central helium abundance is $Y_\mathrm{c}>10^{-4}$)
were computed with mass loss rates prescribed by the Reimers formula (Reimers 1975) with parameter
$\eta_\mathrm{R}=0.5$, whereas evolution of AGB stars was computed according to Bl\"ocker (1995)
with parameter of the Bl\"ocker formula $\eta_\mathrm{B}=0.05$.
Additional calculations with mass loss parameters $\eta_\mathrm{B}=0.03$ and $\eta_\mathrm{B}=0.07$
were carried out for evolutionary sequences that showed a good agreement with observations of R~Hya.
In general we computed several dozen evolutionary sequences and then selected models of these sequences
were used as initial conditions in hydrodynamic calculations for determination of the pulsation period.

\begin{figure*}[p!]
%%% Figure:1
\centerline{\includegraphics[width=0.7\textwidth]{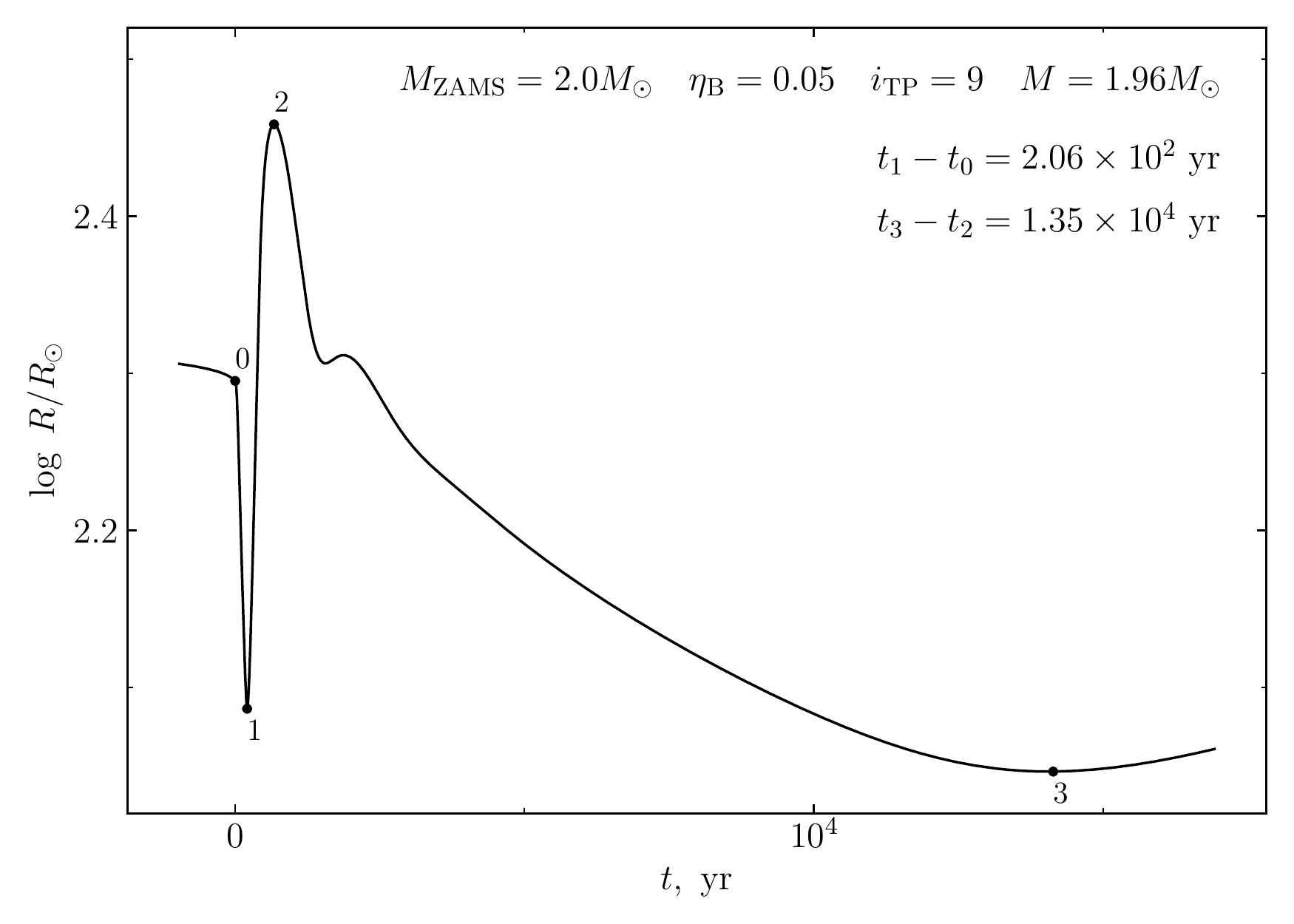}}
\caption{Radius variation of the star with mass $M=1.96M_\odot$ (evolutionary sequence
         $M_\textrm{ZAMS}=2.0 M_\odot$, $\eta_\mathrm{B}=0.05$) during the thermal flash
         $i_\mathrm{TP}=9$.
         Evolutionary time $t$ is set to zero maximum of $L_\mathrm{3\alpha}$. \hfill}
\label{fig1}
\end{figure*}

\begin{figure*}[p!]
%%% Figure:2
\centerline{\includegraphics[width=0.7\textwidth]{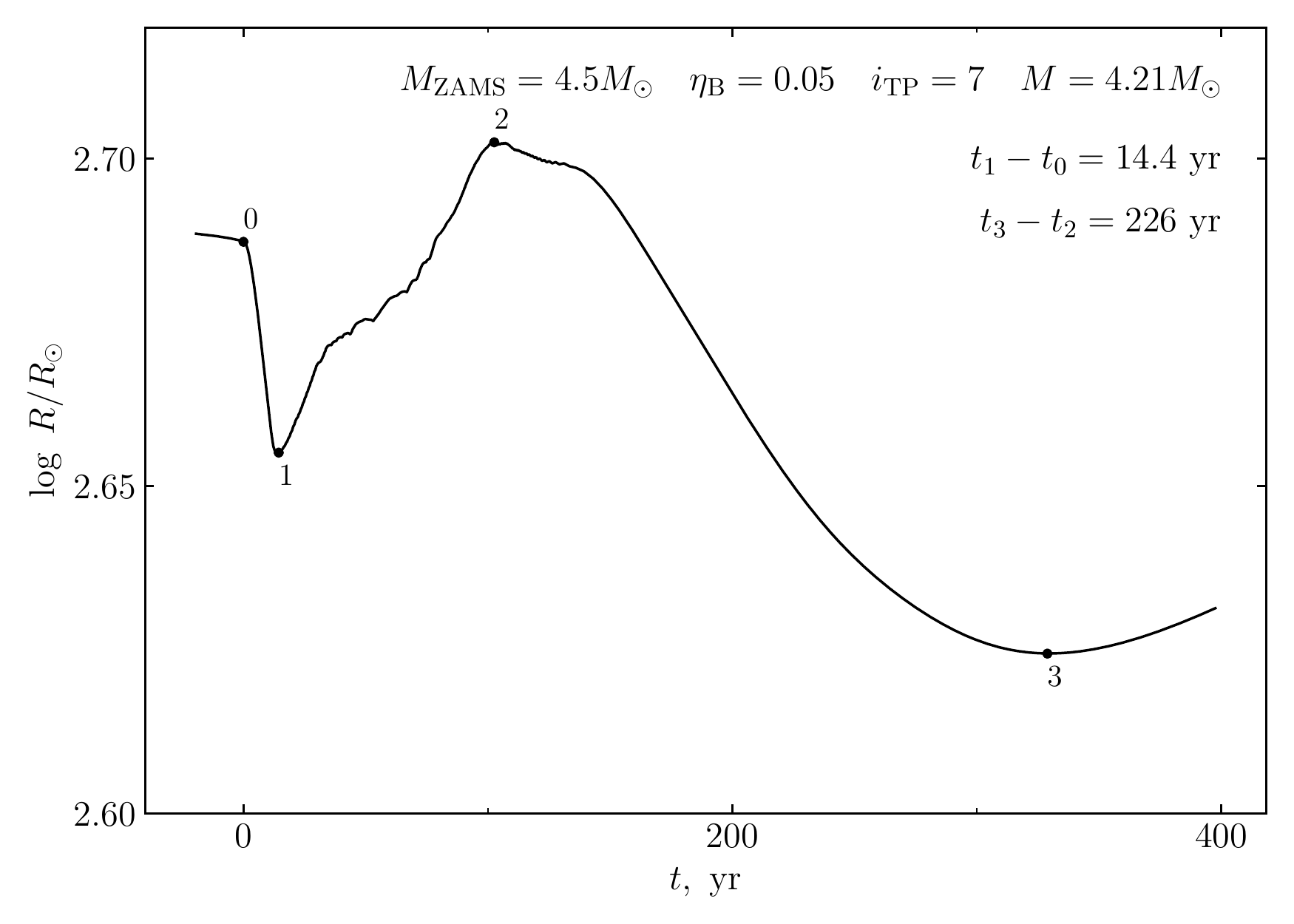}}
\caption{Same as Fig.~1 but for the star with mass $M=4.21M_\odot$
         (evolutionary sequence $M_\textrm{ZAMS}=4.5M_\odot$, $\eta_\mathrm{B}=0.05$) during
         the thermal flash $i_\mathrm{TP}=7$. \hfill}
\label{fig2}
\end{figure*}

The pulsation period and the stellar radius relate as $\Pi\propto R^{3/2}$ so that the temporal
dependence of the pulsation period after the maximum of the helium shell luminosity $L_{3\alpha}$
can be obtained from that of the radius without time consuming hydrodynamic computations.
Fig.~\ref{fig1} shows variation with time during the ninth thermal flash ($i_\mathrm{TP}=9$)
of the stellar radius $R$ in the star with mass $M=1.96M_\odot$
(the evolutionary sequence $M_\textrm{ZAMS}=2M_\odot$, $\eta_\mathrm{B}=0.05$).
A similar plot for the star with mass $M=4.21M_\odot$
(the evolutionary sequence $M_\textrm{ZAMS}=4.5M_\odot$, $\eta_\mathrm{B}=0.05$)
during the seventh thermal flash ($i_\mathrm{TP}=7$)
is shown in Fig.~\ref{fig2}.
In both Figs.~\ref{fig1} and ~\ref{fig2} the time $t$ is set to zero at the maximum of $L_{3\alpha}$
and the maxima of $L_{3\alpha}$ are marked by filled circles with label 0.

\begin{figure*}[t!]
%%% Figure:3
\centerline{\includegraphics[width=0.7\textwidth]{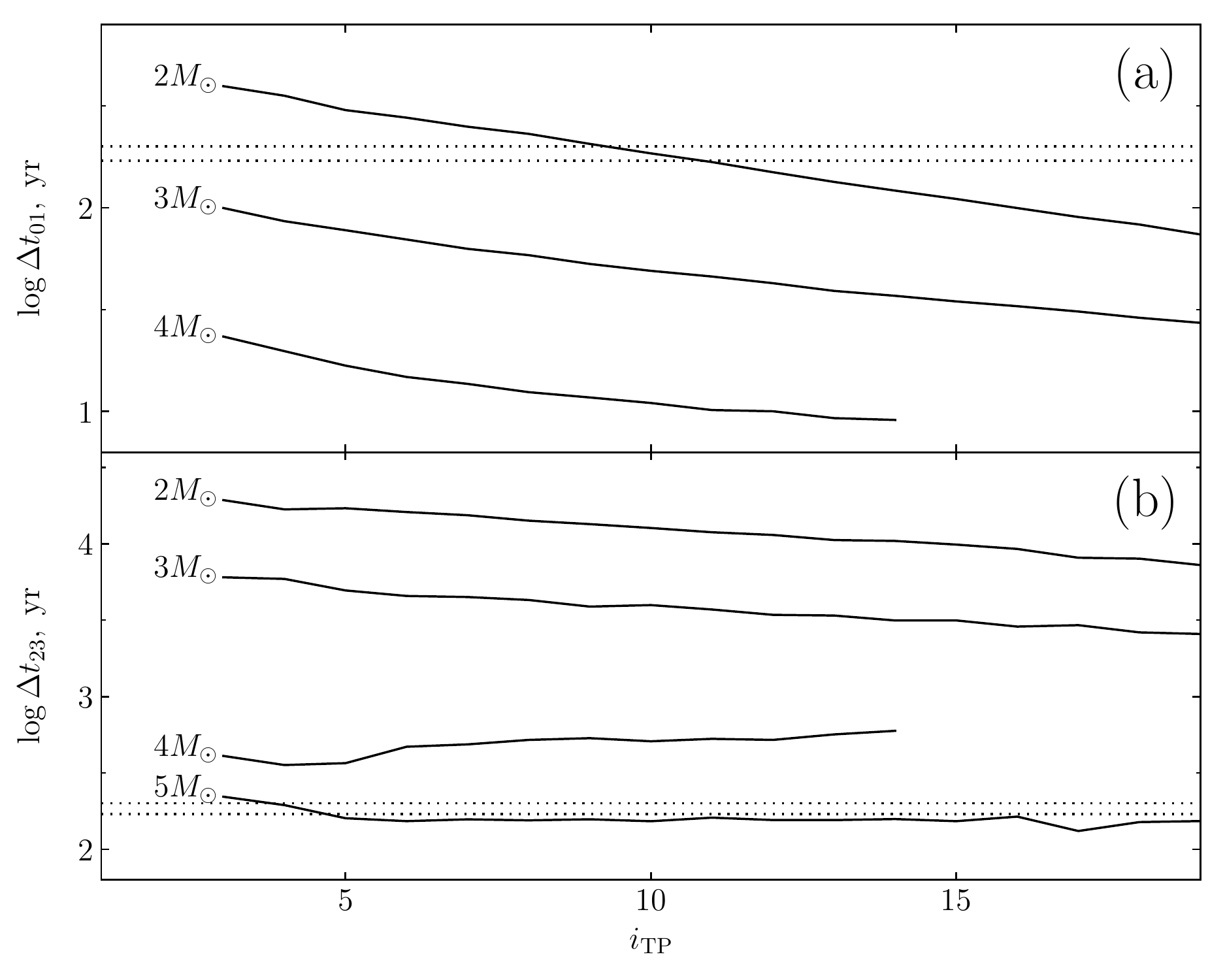}}
\caption{Duration of the first (a) and the second (b) phases of radius decrease as
         a function of the thermal flash serial number $i_\mathrm{TP}$ for
         evolutionary sequences calculated with $\eta_\mathrm{B}=0.05$.
         The initial masses $M_\textrm{ZAMS}$ are given at the curves.
         Dotted lines show the time intervals 170 and 200~yr. \hfill}
\label{fig3}
\end{figure*}

At maximum of $L_{3\alpha}$ the gas in the hydrogen burning shell adiabatically expands
so that the rate of energy release in the CNO cycle abruptly drops whereas outer layers of the star
contract due to the lack of hydrostatic equilibrium.
Decrease of the stellar radius and luminosity ceases at $t_1$ when the radiation diffusion wave
generated by the helium flash reaches the bottom of the outer convection zone.
Therefore, the time interval $\Delta t_{01} = t_1 - t_0$ is nearly the radiation diffusion time
between the helium burning shell and the outer layers of the star.
As seen in Figs.~\ref{fig1} and \ref{fig2} the time interval $\Delta t_{01}$ decreases with increasing
stellar mass.
Duration of the second phase of radius decrease ($\Delta t_{23} = t_3 - t_2$)
exceeds the time interval $\Delta t_{01}$ by more than an order of magnitude but also decreases
with increasing stellar mass.

As seen in Figs.~\ref{fig1} and \ref{fig2}, the time intervals for stellar masses $M=1.96M_\odot$
($\Delta t_{01} = 206$ yr) and $M=4.21M_\odot$ ($\Delta t_{23} = 226$ yr) are nearly the same and
are close to the time interval ($170~\textrm{yr}\lesssim \Delta t\approx 200~\textrm{yr}$)
of period decrease observed in R~Hya.
Therefore, in order to compute the model of the Mira--type variable R~Hya we have to consider
both phases of stellar radius decrease.

\begin{figure*}[t!]
%%% Figure:4
\centerline{\includegraphics[width=0.7\textwidth]{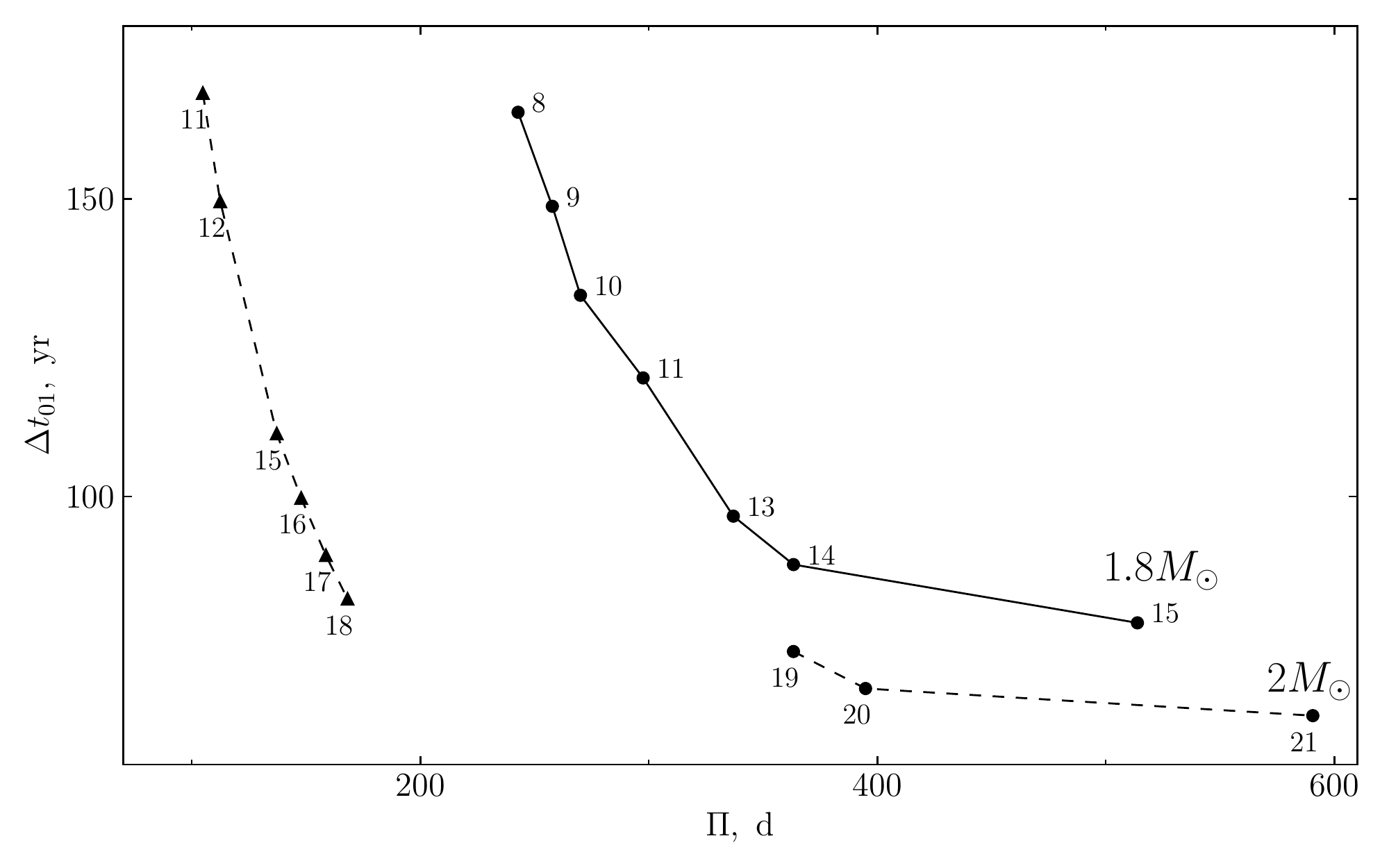}}
\caption{The diagram period $\Pi$~--- duration of the first phase of radius decrease
         $\Delta t_{01}$ for evolutionary sequences $M_\textrm{ZAMS}=1.8M_\odot$ (solid line) and
         $M_\textrm{ZAMS}=2M_\odot$ (dashed line).
         Filled circles and trangles show the fundamental mode and the first overtone pulsators.
         The numbers at the curves show the serial number of the thermal flash $i_\mathrm{TP}$. \hfill}
\label{fig4}
\end{figure*}

The time interval of radius decrease depends not only on the stellar mass but also on the star age
of the AGB star.
This is illustrated in Fig.~\ref{fig3}, where the time intervals $\Delta t_{01}$ and $\Delta t_{23}$ are
shown as a function of the serial number of the thermal flash $i_\mathrm{TP}$ for several evolutionary sequences
with initial masses $2M_\odot\le M_\textrm{ZAMS}\le 5M_\odot$.
As seen in the plots, the time interval $\Delta t_{01}$ is rather close to the observed value for
the early AGB stage ($i_\mathrm{TP} < 10$) in evolutionary sequences with initial masses $M_\textrm{ZAMS} < 3M_\odot$.
The duration of the second phase of radius decrease $\Delta t_{23}$ is close to observations in
wider ranges of the serial number of the thermal flash ($i_\mathrm{TP} > 5$) but for evolutionary sequences with
higher initial masses: $4M_\odot < M_\textrm{ZAMS} \lesssim 5M_\odot$.
To draw a more certain conclusion about applicability of the first or the second phase of radius
decrease we have to evaluate the pulsation periods $\Pi_0$ and $\Pi_2$ at the onset of the radius diminution.

\section*{hydrodynamic pulsation models of red giants}

Pulsation periods of the Mira--type star models were determined using the discrete Fourier transform
of the kinetic energy of the limit--cycle stellar pulsation motions.
The self--excited nonlinear stellar pulsations with transition to the limit--cycle oscillations were
obtained as the solution of the Cauchy problem for equations of radiation hydrodynamics with initial
conditions determined from selected hydrostatically equilibrium evolutionary models.
Effects of time--dependent convection were taken into account by solution of the transport equations for
diffusion of the specific enthalpy and the mean kinetic energy of turbulent motions (Kuhfu\ss (1986).
Basic equations of hydrodynamics and parameters of the theory of time--dependent convection are described
in (Fadeyev 2013).

\begin{figure*}[t!]
%%% Figure:5
\centerline{\includegraphics[width=0.7\textwidth]{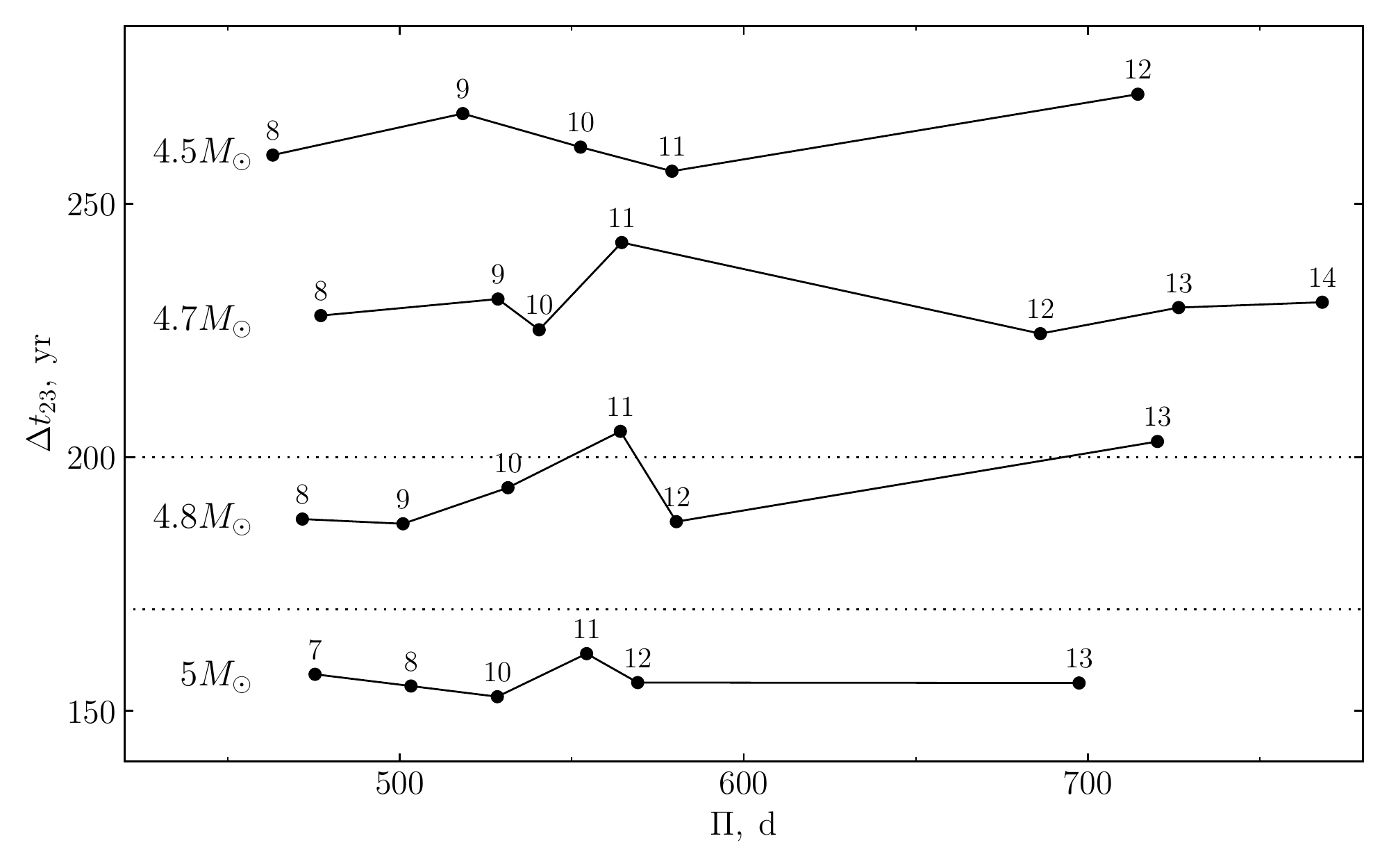}}
\caption{The diagram period $\Pi$~--- duration of the second phase of radius decrease
         $\Delta t_{23}$ for evolutionary sequences $M_\textrm{ZAMS}=4.5M_\odot$,
         $4.7M_\odot$, $4.8M_\odot$ and $5M_\odot$ computed for $\eta_\mathrm{B}=0.05$.
         Dotted lines show the time interval
         $170~\textrm{yr} \leq\Delta t \leq 200~\textrm{yr}$. \hfill}
\label{fig5}
\end{figure*}

A substantial difficulty encountered in pulsation calculations of Mira--type stars is due to the fact
that the hydrostatically equilibrium stellar envelope might be in thermal imbalance whereas the thermal
equilibrium is the necessary condition for application of the theory of stellar pulsation and for correct
evaluation of the pulsation period (Ya'Ari and Tuchman 1996).
To avoid this obstacle we used an approach based on evaluation of the degree of deviation from
thermal equilibrium within the stellar envelope (Fadeyev 2022).
To this end we evaluated the quantity
$$
\delta_\mathrm{L} = \max_{1\le j\le N} |1 - L_j/L_1| ,
$$
where $L_j$ is the radiative plus convective luminosity at the $j$--th mass zone of the
hydrodynamic model.
The inner boundary ($j=1$) is treated as a rigid permanently radiating sphere and at the outer boundary
$j=N$.
All hydrodynamic computations were carried out with $N=600$ mass zones.
The condition $\delta_\mathrm{L}=0$ obviously corresponds to the thermal equilibrium.
The criterion of enough small deviations from thermal equilibrium when the theory of stellar
pulsation is applicable is given by the condition $\delta_\mathrm{L} \lesssim 10^{-2}$
(Fadeyev 2022).
Computations carried out in the present study showed that this condition is fulfilled
in vicinity of stellar radius extremums shown in Figs.~\ref{fig1} and \ref{fig2} by filled circles.

The relationship between the duration of the first phase of radius decrease $\Delta t_{01}$ and the
pulsation period $\Pi$ at the maximum of $L_{3\alpha}$ is shown in Fig.~\ref{fig4} for evolutionary
sequences $M_\textrm{ZAMS}=1.8M_\odot$ and $M_\textrm{ZAMS}=2M_\odot$ computed with $\eta_\mathrm{B}=0.05$.
Increase of pulsation period $\Pi$ with increasing serial number of the thermal flash $i_\mathrm{TP}$ is due
to the fact that the stellar luminosity and the radius at the maximum of $L_{3\alpha}$ gradually grow
as the AGB star evolves.
The time interval $\Delta t_{01}$ gradually decreases with increasing $i_\mathrm{TP}$ and, therefore, with
increasing pulsation period.
As can be seen in Fig.~\ref{fig4}, the pulsation periods of red giants with masses $M\lesssim 2M_\odot$
at the maximum of $L_{3\alpha}$ in general do not exceed 400 d and radial pulsations with periods
$\Pi\approx 500$ d exist only during the final stage of AGB when the time interval $\Delta t_{01}$
is less than 100 yr.
Therefore, the assumption that the secular period reduction of R~Hya takes place during the first phase
of radius decrease should undoubtedly be rejected.

The diagram illustrating the relationship between the pulsation period and the time interval
$\Delta t_{23}$ shown in Fig.~\ref{fig5} allows us to conclude that an acceptable agreement
between the theory and observations is obtained for the evolutionary sequence $M_\textrm{ZAMS}=4.8M_\odot$.
It should be noted that the non--monotonic change in time interval $\Delta t_{23}$ with increasing
$i_\mathrm{TP}$ is due to irregular variations of the maximum radius $R_2$.

\begin{table}
\caption{Models of the Mira--type star R~Hya with initial mass $M_\textrm{ZAMS}=4.8M_\odot$}
\label{tabl1}
\begin{center}

\begin{tabular}{c|c|c|c|c|c|c|c|c}
\hline
$\eta_\mathrm{B}$ & $i_\mathrm{TP}$ & $M/M_\odot$ & $X_\mathrm{C}/X_\mathrm{O}$ & $\Delta t_{23},\ \textrm{yr}$ & $R_2/R_\odot$ & $R_3/R_\odot$ & $\Pi_2,\ \textrm{d}$ & $\Pi_3,\ \textrm{d}$  \\
\hline
 0.03 &   6 &   4.63 &  0.305 &    177 &    495 &    429 &    451 &    383 \\
      &   7 &   4.60 &  0.331 &    164 &    512 &    437 &    469 &    387 \\
      &   8 &   4.58 &  0.355 &    154 &    532 &    445 &    473 &    393 \\
 0.05 &   5 &   4.58 &  0.279 &    197 &    498 &    421 &    470 &    377 \\
      &   6 &   4.54 &  0.302 &    173 &    502 &    429 &    473 &    389 \\
      &   7 &   4.50 &  0.328 &    189 &    515 &    435 &    471 &    394 \\
 0.07 &   5 &   4.50 &  0.282 &    200 &    499 &    424 &    472 &    386 \\
      &   6 &   4.44 &  0.306 &    177 &    506 &    433 &    463 &    398 \\
\hline
\end{tabular}

\end{center}
\end{table}

\section*{models of the mira--type variable r hya}

Early observational estimates of the pulsation period of R~Hya $\Pi_a\approx 495$ d
obtained at the turn of XVIIth--XVIIIth centuries are scarce and rather unreliable
(Zijlstra et al. 2002), so that computation of R~Hya model should take into account
the period value $\Pi_b\approx 380$ d corresponding to nearly 1950 when the period
decrease ceased.
To this end we computed additional evolutionary sequences of AGB stars with initial mass
$M_\textrm{ZAMS}=4.8M_\odot$ for the mass loss parameters $\eta_\mathrm{B}=0.03$ and
$\eta_{\textrm{B}}0.07$.
Selected models of these sequences corresponding to the second radius maximum $R_2$
and the second radius minimum $R_3$ were used as initial conditions in hydrodynamic
computations and finally for determination of pulsation periods $\Pi_2$ and $\Pi_3$.

Results of these computations are collected in table~\ref{tabl1} for models with
initial mass $M_\textrm{ZAMS}=4.8M_\odot$, where in first four columns we give the mass loss
parameter $\eta_\mathrm{B}$ (Bl\"ocker 1995), the serial number of the thermal flash $i_\mathrm{TP}$,
the mass of the star at maximum of $L_{3\alpha}$ (i.e. for t=0)
and the ratio of carbon ${}^{12}\textrm{C}$ to oxygen ${}^{16}\textrm{O}$ mass
abundances at the outer boundary of the evolutionary model.
It should be noted that at the early AGB stage prior to the first thermal pulse
the surface ratio of mass abundances is $X_\mathrm{C}/X_\mathrm{O}=0.248$ and
therefore the composition of models listed in table~\ref{tabl1} is characterized
by the enhanced carbon abundance indicating the stage of the 3rd dredge--up.
At the same time all the models remain the oxygen--rich Miras because the ratio
of surface concentrations is $N_\mathrm{C}/N_\mathrm{O} < 1$.

All models listed in table~\ref{tabl1} satisfy the condition that the pulsation
period $\Pi_3$ differ from the observational estimate $\Pi_b^\star=380$ d
by less than 5\%.

\section*{conclusions}

Results of stellar evolution and nonlinear stellar pulsation calculations allow us
to draw an unambiguous conclusion that the period decrease observed in R~Hya 
corresponds to the second phase of radius decrease when the radiation--diffusion wave
from the helium burning shell dissipates in the outer layers of the star.
Earlier the same conclusion was drawn by Wood and Zarro (1981) on the basis of their
evolutionary calculations.
At the same time we have to note the significant difference between estimates of
stellar parameters presented by Wood and Zarro (1981) and those obtained in our study.
In particular, the mass of the carbon--oxygen degenerate core and the stellar luminosity
of our model are $M_\mathrm{CO}=0.856M_\odot$ and $L\approx 2.5\times 10^4 L_\odot$, whereas estimates
obtained by Wood and Zarro (1981) are significantly smaller: $M_\mathrm{CO}=0.653M_\odot$,
$L\approx 1.3\times 10^4 L_\odot$.
This disagreement is due to the fact that estimates by Wood and Zarro (1981) were obtained
for the significantly longer phase of radius decrease and did not take into account the fact
that the period of R~Hya has ceased to decrease in $\approx 1950$.

R~Hya is one of the nearby long--period pulsating variables and by now the
measurements of its angular diameter were carried out with methods of the optical
interferometry.
According to Haniff et al. (1995) the angular diameter of R~Hya is $d=33$ mas and
for the distance 125 pc its stellar radius is $R=442R_\odot$.
It should be noted that this distance estimate was obtained from the approximate
period--luminosity--color relation (Feast et al. 1989).
However the recent distance estimate based on the astrometric catalog Gaia DR3 is 126 pc
(Andriantsaralaza et al. 2022) so that the observational estimate of the radius remains
the same.
Thus agreement of theoretical estimates of radii $R_3$ listed in table~\ref{tabl1} with
observations confirm the theoretical estimates of the stellar mass:
$4.44M_\odot\le M\le 4.63M_\odot$.
The range of mass estimates is less than 5\%.
This is not only due to variations of mass loss rate parameter $\eta_\mathrm{B}$, but also due to
computational accuracy limitations hampering a reliable evaluation of the maximum stellar
radius $R_2$.

\section*{references}

\begin{enumerate}

\item M. Andriantsaralaza, S. Ramstedt, W.H.T. Vlemmings, and E. De Beck,
      Astron. Astrophys. \textbf{667}, A74 (2022).

\item M. Asplund, N. Grevesse, A.J. Sauval, and P. Scott,
      Annual Rev. Astron. Astrophys. \textbf{47}, 481 (2009).

\item T. Bl\"ocker, Astron. Astrophys. \textbf{297}, 727 (1995).

\item E. B\"ohm--Vitense, Zeitschrift f\"ur Astrophys. \textbf{46}, 108 (1958).

\item A.J. Cannon and E.C. Pickering, Annals of Harvard College Observatory \textbf{55}, 95 (1909).

\item S. C. Chandler, Astron. Nachr. \textbf{103}, 225 (1882).

\item R.H. Cyburt, A.M. Amthor, R. Ferguson, Z. Meisel, K. Smith, S. Warren, A. Heger, R.D. Hoffman,
      T. Rauscher, A. Sakharuk, H. Schatz, F.K. Thielemann, and M. Wiescher,
      Astrophys. J. Suppl. Ser. \textbf{189}, 240 (2010).

\item Yu.A. Fadeyev, Astron. Lett. \textbf{39}, 306 (2013).

\item Yu.A. Fadeyev, MNRAS \textbf{514}, 5996 (2022).

\item M.W. Feast, I.S. Glass, P.A. Whitelock, and R.M. Catchpole, MNRAS \textbf{241}, 375 (1989).

\item B.A. Gould, Astron. Nachr. \textbf{102}, 341 (1882).

\item C.A. Haniff, M. Scholz, and P.G. Tuthill, MNRAS, \textbf{276}, 640 (1995).

\item F. Herwig, Astron. Astrophys. \textbf{360}, 952 (2000).

\item F. Herwig, N. Langer, and M. Lugaro, Astrophys. J. \text{593}, 1056 (2003).

\item D. Hoffleit, J. Am. Associat. Var. Star Observ. \textbf{25}, 115 (1997).

\item E.M.L. Humphreys, M.D. Gray, J.A. Yates, and D. Field, MNRAS \textbf{287}, 663 (1997).

\item M.J. Ireland, P.G. Tuthill,  T.R. Bedding, J.G. Robertson, and A.P. Jacob,
      MNRAS \textbf{350}, 365 (2004).

\item R. Kuhfu\ss, Astron. Astrophys. 160, 116 (1986).

\item T.Lebzelter and J. Hron, Astron. Astrophys. \textbf{411}, 533 (2003).

\item B.M. Lewis, P. David P,  and A.M. Le Squeren, Astron. Astrophys. Suppl. Ser. \textbf{111}, 237 (1995).

\item S.J. Little, I.R. Little--Marenin, and W.H. Bauer,  Astron. J. \textbf{94}, 981 (1987).

\item H. Ludendorff, Astron. Nachr. \textbf{203}, 117 (1916).

\item H. Maehara, Publ. Astron. Soc. Japan \textbf{23}, 313 (1971).

\item P.W. Merrill,  Astrophys. J. \textbf{103}, 6 (1946).

\item P.W. Merrill, Publ. Astron. Soc. Pacific \textbf{69}, 77 (1957).

\item R. M\"uller, Astron. Nachr. \textbf{237}, 81 (1929).

\item A.V. Nielsen, Astron. Nachr. \textbf{227}, 141 (1926).

\item M.Ya. Orlov and A.V. Shavrina, Nauchn. Inform. Astron. Sovet AN SSSR \textbf{56}, 97 (1984).

\item B. Paxton, R. Smolec, J. Schwab, A. Gautschy, L. Bildsten, M. Cantiello, A. Dotter,
      R. Farmer, J.A. Goldberg, A.S. Jermyn, S.M. Kanbur, P. Marchant, A. Thoul, R.H.D. Townsend, W.M. Wolf,
      M. Zhang, and F.X. Timmes, Astrophys. J. Suppl. Ser. \textbf{243}, 10 (2019).

\item M. Pignatari, F. Herwig, R. Hirschi, M. Bennett, G. Rockefeller, C. Fryer, F.X. Timmes,
      C. Ritter, A. Heger, S. Jones, U. Battino, A. Dotter, R. Trappitsch, S. Diehl, U. Frischknecht, A. Hungerford,
      G. Magkotsios, C. Travaglio, and P. Young, Astrophys. J. Suppl. Ser. \textbf{225}, 24 (2016).

\item D. Reimers, \textit{Problems in stellar atmospheres and envelopes} (Ed. B. Baschek,
      W.H. Kegel, G. Traving, New York: Springer-Verlag, 1975), p. 229.

\item N.N. Samus', E.V. Kazarovets, O.V. Durlevich, N.N. Kireeva, and E.N. Pastukhova,
      Astron. Rep. \textbf{61}, 80 (2017).

\item J.F.J. Schmidt, Astron. Nachr. \textbf{65}, 173 (1865).

\item H. Takaba, I. Takahiro, M. Takeshi, and S. Deguchi, Publ. Astron. Soc. Japan \textbf{53}, 517 (2001).

\item P.R. Wood and D.M. Zarro, Astrophys. J. \textbf{247}, 247 (1981).

\item H.C. Woodruff, P.G. Tuthill, J.D. Monnier, M.J. Ireland, T.R. Bedding, S. Lacour,
      W.C. Danchi, and M. Scholz,
      Astropjys. J. \textbf{673}, 418 (2008).

\item A. Ya'Ari and Y. Tuchman, Astrophys. J. \textbf{456} 350 (1996).

\item A.A. Zijlstra, T.R Bedding, and J.A. Mattei, MNRAS \textbf{334}, 498 (2002).
\end{enumerate}

\end{document}